# Nano-molar to milli-molar level Ag (I) determination using absorption of light by ZnS QDs without organic ligand

Rabindra Nath Juine[a], S. Amirthapandian[b], S. Dhara[c], A. Das[c*].

[a] Health Physics Unit, Nuclear Recycle Board, Bhabha Atomic Research Centre Facilities, Kalpakkam 603102, India. [b] Materials Physics Division, Materials Science Group, Indira Gandhi Centre for Atomic Research, HBNI, Kalpakkam 603102, India. [c] Surface and Nanoscience Division, Indira Gandhi Centre for Atomic Research, HBNI, Kalpakkam 603102, India.

* Corresponding author: Dr. A. Das, Tel: +044-274-80500 (22409). Email: dasa@igcar.gov.in

**Abstract**

Despite of useful applications of Ag, it is hazardous to health and environment and hence early detection is required. Crystal defects influencing optical property is less likely utilized for Ag detection using prevalent UV-Vis technique. Quantum dots (QDs) ZnS prepared by soft chemical route are exploited for the detection of $Ag^+$ in aqueous solution for nM to mM concentrations. UV-Vis and photoluminescence (PL) measurements reveal quantum confinement effect and presence of defects in ZnS nanoparticles (NPs). Controlled synthesis of ZnS NPs allows appearance of defects related peak at 400–550 nm in UV-Vis spectra. A plausible mechanism is presented and elucidated by XRD and PL studies for chemical interaction between ZnS with $Ag^+$. The interaction affects prominently the defects related absorption of ZnS NPs and allows single step $Ag^+$ detection in aqueous medium. This method is extended to other ions including 'S' prone heavy and toxic ions like Pb, Hg, and Cd. Among them, $Ag^+$ shows the best response with ZnS NPs. This demonstration using a portable and cost effective technique with no organic component opens up possibility for other inorganic materials.

1. Introduction

Noble metal Ag finds wide spread applications as excellent electrical, thermal conductors and metal reflectors including in plasmonic, anti-bacterial uses, and jewel industries [1-4]. Use of Ag as an antimicrobial agent has been increased in medical devices for instances, in catheters, endotracheal breathing tubes, bone prostheses [3, 4]. Ag is also well utilized as wound dressing and as colloidal solution for health benefits [2-5]. Importantly, Ag NPs with large specific surfaces being more effective than the colloidal Ag particles [6, 7] find ever increasing applications in consumer-oriented products e.g. in detergents, clothing, and food additives [8]. Despite of its large scale useful applications, Ag is also found to be a hazardous material and potent threat to health and environment [6]. In fact, Ag ions are highly toxic to aquatic life and affect permeability of cell membrane by disrupting $K^+$ concentration [1, 9, 10]. Owing to strong oxidation nature, it can lead to internal organ edema and even to death [11] and chronic exposure to Ag leads to argyria [8]. The Environmental Protection Agency in U.S. has set a secondary maximum contaminant level (SMCL) for Ag at 0.1 mg/L (0.93 µM) [12]. In toxic argyrosis, concentration of $Ag^+$ in serum and urine goes up to a value of 2 µM [2]. It is reported that Ag NPs releases $Ag^+$ for effectiveness as antibacterial applications [7]. The accurate and convenient determination of Ag ions is thus a demanding challenge. Moreover, trace quantity determination (<µM) can allow early detection. Common techniques for the Ag ions determination are atomic absorption spectrometry, plasma emission spectrometry, anodic stripping voltammetry [5], electrochemical probe [11], colorimetric sensing probe [7, 13] fluorescent probe [14-20], photo luminescent intensity variation [21, 22] and chemosensor [23-25]. Algarra *et. al.* [14] demonstrated $Ag^+$ detection based on change in fluorescent intensity using carbon dots which was tailored with thiols followed by binding with mercaptosuceinic acid to its surface. Their detection method was employed for the range of nM. Similarly, Gao *et. al.* [15] utilized luminescent carbon-dots to detect Ag clusters in water medium linearly for the range of 0-90 µM with a limit of detection to 320 nM. Hao *et. al.* [16] detected $Ag^+$ in aqueous solutions using 3D micro porous compound which acted as a fluorescence probe with a detection limit up to 0.1 µM. Similarly, chemically active naphthalimide derivative was used for studying variation of fluorescence intensity for detection of $Ag^+$ [17]. Determination $Ag^+$ was also reported using organic ligand with ZnS NPs (e.g. Thiolactic acid-ZnS) [18] where this specific chemical compound acted as fluorescence probe. It allowed observation of sensitive quenching mainly in the range of 0–0.5 μM. Wen *et. al*. reported quenching effects [19] that provided $Ag^+$ determination down to nM using graphene oxide and cytosine-rich oligonucleotide as fluorogenic probe. Whereas Hatai *et. al.* [23] developed chemosensor for $Ag^+$ using a bis-thiocarbamate scaffold which displayed fluorescence enhancement in the presence of $Ag^+$ and the detection limit was reported to be as low as 11 ppb. Recently, Zhao *et. al*. [26] have utilized specific graphene QDs as a reference fluorophore along with another organic material, *o*-phenylenediamine to probe $Ag^+$ in aqueous medium by fluorescence resonance energy transfer (FRET) process. All above methods detected very low quantity of $Ag^+$. However, they have used organic capping ligand or coating and multi-step process for optical sensing. This added compound is found to influence sensitivity with reference to pH [14, 24] or treated medium [21]. The low detection limit is thus linked to utilization of special organic chemicals which increase the expenses and preparation time. Hence, there is outlook for developing simple and cost effective alternative method. UV-Vis spectroscopy, a simple technique was used earlier with a chemically functionalized material which showed probe peak for detection of Ag [5, 23, 27]. Interestingly, NPs containing sulfur like ZnS can be a source of noble metal (Ag, Au, Cu) -S interaction [27-29]. Any technique responding to this bond formation can allow recognition of $Ag^+$. Importantly, use of UV-Vis spectroscopy will be a cost-effective and fast method. However there is little information on direct detection of $Ag^+$ using inorganic NPs. In the present investigation, strong bonding affinity of sulfur to noble metals, in particular Ag by forming Ag-S bond [30, 31] is utilized. In this context, the point of focus is huge reactive surfaces of ZnS NPs, in particular QD regime for the detection of trace quantity. In our earlier report [32], QD ZnS has shown presence of various defects including elemental type of sulfur defects at surfaces. Such suitable surface defects can act as interaction sites for $Ag^+$.

ZnS is an *n*-type semiconductor with a high optical band gap of 3.67 eV. It acts as an important fluorescence material [33] and is potent biocompatible fluorescence for detection of maligned tissues [34]. They are found in Zinc Blend (ZB) and Wurtzite crystalline structures. The former is most stable structure in bulk form. However nanostructured ZB system can have both threefold and two fold coordinated atoms at the edges [35]. For atoms at edges, ZB-ZnS nanostructure has greater surface energy than that in the wurtzite-ZnS nanostructures. It can render surfaces of ZB nanostructure chemically reactive for interaction, especially with sulfur prone metals at room temperature. However, this chemical interaction is not used with sulfur containing NPs like ZnS pertinent to detection of hazardous metals.

In the present report above conjecture is discussed by synthesizing different sizes of ZB ZnS NPs which offer by far the largest possible surfaces in QDs ZnS. Cost effective portable fiber optic based UV-Vis spectroscopy is employed for monitoring absorption of ZnS for detection method. Pertaining to sizes of ZnS NPs, different amounts of surfaces defects arise and they actively control surface induced reactions for detection of $Ag^+$. This approach utilizes defects in ZnS NP appearing in visible range allowing wide range of detection starting from milli-molar (mM) to nano-molar (10 nM) of $Ag^+$. Further investigation unraveling mechanism of detection related to ZnS - $Ag^+$ interaction is also carried out. Same defects peak of ZnS is used to understand reaction mechanism for 'S' prone heavy elements e.g. Pb, Hg and Cd. Other alkali and alkaline ions are also tested with ZnS NPs and compared for detection responses with $Ag^+$ which shows strongest response.

2. **Experimental**

Zinc Sulfide (ZnS) was prepared by wet chemical method using zinc chloride, ($ZnCl_2$, MERCK) and sodium sulfide ($Na_2S$, SD Fine chemicals, India) as starting material. Three different concentrations of 0.1 wt% (7.35 mM), 0.25 wt%, and 0.5 wt% of $ZnCl_2$ were used to synthesize different sizes of ZnS NPs with $Na_2S$ (0.03 M). Detailed description was given elsewhere [32]. In brief, $ZnCl_2$ was dissolved in MilliQ water (18.2 M$\Omega$.cm) and the solution was stirred magnetically for 30 min at room temperature. Similarly, 1 wt% $Na_2S$ (0.03 M) solution was prepared and was added drop wise to solution of $ZnCl_2$ at elevated temperature of 45 ± 2 °C at atmospheric condition. The resultant white precipitate was collected by centrifuge at 10000 rpm. The precipitate was rinsed several times to remove any unreached precursors e.g. chloride and sulfides with MilliQ water. Collected precipitates were dried and grinded. Similarly ZnS NPs were made for all $ZnCl_2$ concentrations. Structural characterization was carried out using X-ray diffraction (Inel EQUINOX 2000) measurements of powder ZnS NPs with Cu $K_\alpha$ radiation. High resolution transmission electron microscopy (HRTEM, LIBRA 200FE) was used for structural and morphological studies of ZnS NPs. A low concentration of NPs (0.5 mg/2 ml) was dispersed and ultrasonicated in isopropanol for 5 min. A few drops of it was placed on the carbon coated

Cu-grid for HRTEM studies. For recording the UV-Vis absorption spectra, a suspension of ZnS NPs (0.5 mg/2 ml) in distilled water was used. Spectrometer records data for the range of interest (225-750 nm) within 20 ms/cycle and its noise level is better than 0.1% absorption scale. This portable system uses fiber optics with a high resolution AvaSpec 3648 USB detector. Absorption spectrum provided the absorption co-efficient ($\alpha$) and the optical band gap of the material using Tauc's Plot [32] The PL measurements were carried out using 325 nm excitation from He-Cd laser (InVia; Reinshaw).

### 2.1. Experiment for $Ag^+$ detection

Different concentrations of $AgNO_3$ solution prepared for the source of $Ag^+$ were kept in dark place to protect light exposure. Firstly, absorption spectrum of known amount of ZnS NPs in water (3 ml) is recorder against reference water with 50 cycles for improved data where each cycle is recorded in 20 ms. It serves as staring absorption spectrum of ZnS NPs for detection of $Ag^+$. Now different aliquot of $AgNO_3$ solution was added to same concentration of (3ml) ZnS NPs solution and subsequent change for a particular concentration of $Ag^+$ is recorded immediately after addition. This process was carried out for the all three sizes of ZnS NPs. Similar to $AgNO_3$ solution, other metal ions solutions were also prepared for Na, K, Mg, Ca, Ba, Mn, Hg, Pb, and Cd using their soluble salts and are used separately for reaction with ZnS NPs. Noteworthy, commonly found ions in aqueous medium are various salt of alkali and alkaline metals. This study helps understanding mechanism as well as response for various metal ions in aqueous medium.

### 3. Result and discussion
### 3.1. Characterization of ZnS NPs

Structural and morphological details of ZnS NPs were obtained by using X-ray diffraction (XRD) and HRTEM. XRD patterns of different sizes of ZnS NPs are shown in Fig. 1. Five strong and broad peaks are observed for all of them. These peaks correspond to (111), (220), (311), (400) and (331) planes of zinc blend structure of ZnS (JCPDS # 05-0566). Noticeably, a systematic decrease was found in peaks width with increasing concentration of $ZnCl_2$ (Fig. 1). Particle size was estimated for all three ZnS NPs following Debye-Scherrer formula [36] and was found to be $3.0 \pm 0.1$ nm, $3.5 \pm 0.1$ nm, and $4.5 \pm 0.1$ nm for $ZnCl_2$ concentrations of 0.1 wt%, 0.25 wt% and 0.5 wt%, respectively. Thus XRD study confirms preparation of ultrafine zinc blend ZnS in a single step.

Typical HRTEM result for 0.1 wt% and 0.5 wt% samples are shown in Fig. 2a and Fig. 2b, respectively. The as-prepared sample showed nearly spherical particle (Fig. 2a (i) & Fig. 2b (i)) with a lattice spacing, $d$ of 0.31 nm (inset in Fig. 2a (ii) & Fig. 2b (ii)) which matched closely to the $d$ spacing of (111) plane of cubic ZnS phase (JCPDS # 05-0566). Distribution of particle sizes is displayed in Fig. 2a (iii) & Fig. 2b (iii) for 0.1 wt% and 0.5 wt% samples, respectively. Sizes as seen in HRTEM images match to the XRD analysis. Selected area electron diffraction (SAED) displays spotty patterns (Fig. 2a (iv) & Fig. 2b (iv)) due to the presence of crystalline NPs of different orientations. It reveals presence of (111), (220) and (311) planes of cubic ZnS (JCPDS # 05-0566). Similarly HRTEM analysis for 0.25 wt% sample of ZnS showed spherical particles with a maximum size distribution density at 3.7 nm. Details of particle sizes obtained from HRTEM images are given in Table 1.

Above structural and morphological analysis also substantiate optical confinement effect considering Bohr exciton radius (2.5 nm) of ZnS [37]. Absorption spectra were recorded and subsequently, Tauc's plots were deduced for optical band gap energy of three different ZnS NPs (Fig. 3). ZnS being a direct band gap semiconductor, Tauc's plot is given by equation 1 where $\alpha$ is the absorption coefficient and $E_g$ is the bulk band gap energy [39].

$$(\alpha h\nu)^2 = (h\nu - E_g) \quad\ldots\ldots\ldots\ldots\ldots\ldots\ldots\ldots\ldots\ldots\ldots\ldots\ldots (1)$$

Details of estimated optical band gaps are given in Table 1. Quasi quantum confinement was reported even for a size of ~ 5 nm ZnS [38]. Indeed, quasi-quantum confinement effect was clearly seen for particles of 3 nm and 3.5 nm (Table 1) while the band gap was found to increase to a maximum value of 4.5 eV for 3 nm ZnS. However, 4.5 nm ZnS particle grown using 0.5 wt% $ZnCl_2$ did not show blue shift from the reported bulk value of 3.7 eV [38]. Instead, it has slightly lower band gap value of 3.6 ± 0.05 eV. This shift may be due to the presence of large surface defects in finite size ZnS which can influence the band structures.

The shift of band gap is also related to the crystallite size by the relation [39] given below

$$E_g^{eff} = E_g + \frac{h^2\pi^2}{2\mu r^2} - \frac{1.8e^2}{4\pi\varepsilon_0\varepsilon r} \quad\ldots\ldots\ldots\ldots\ldots\ldots\ldots\ldots\ldots\ldots (2)$$

where $E_g^{eff}$ is the observed energy gap of NPs, $E_g$ stands for bulk ZnS value, and $\mu$ is the reduced mass [39]. The calculated particles size is given in Table 1. These particle sizes are found to be comparable to the crystallite sizes obtained from the XRD and HRTEM measurements.

In addition to absorption related to the band gap of ZnS NPs, a prominent broad peak was seen in the range of 400 to 550 nm (Fig. 4). Similar type of absorption peak was reported for ZnS NPs [42]. This unique absorption arises due to large amount of defects in the finite sized ZnS particles, as it does not occur in the bulk ZnS. Moreover, a noticeable variation in the relative absorption among the particles was also observed (Fig. 4). The strong absorption was recorded for the smallest particle which offered the largest surface area and also the large amount of surface related defects as detailed in PL investigation.

PL spectra of ZnS NPs recorded at room temperature are displayed in Fig. 5. Four prominent PL peaks of ZnS NPs were observed at 2.3 (540 nm), 2.55 (485 nm), 2.75 (450 nm), and 2.97 eV (418 nm). They appear due to various defects and closely match with earlier reports for the ZnS nano-materials [35, 40, 41]. At large, these peak positions overlap to the broad features observed in the absorption spectra (Fig. 4). PL peak at 2.3 eV corroborates to elemental type sulfur species on the surface of ZnS NPs [35]. Other two dominant peaks at 2.55 eV and 2.75 eV are assigned to Zn type vacancies [40, 35]. The PL transition was reported to take place as a result of recombination of electron of the sulfur vacancy with trapped holes at Zn acceptor vacancy. Low intense peak for elemental type sulfur was found for larger particles (Fig. 5). Similarly, peak at 2.97 eV was attributed to 'S' vacancy in ZnS nano-structure [40, 35]. For the occurrence of S mediated reactions [30, 31], changes in the peak intensity or position was monitored to measure the reaction kinetics. In that context, PL spectrum was utilized as a quenching probe [18]. Manipulation of defects peaks arising from inorganic NPs as absorption probe for quick detection of hazardous metal interaction is not so far exploited. This lack of exploitation is partly related to difficulty in achieving the process control which can allow routine observation of the absorption peak. It is noteworthy that such absorption is strong function of size and relative amount of defects on the surfaces. Moreover, the related absorption should be stable under measurement conditions. In the present report, we have achieved the absorption peak repeatedly for all particles. It varies, however, according to variation in size of ZnS NPs.

### 3.2. Detection of $Ag^+$ concentration

After the control of the process parameters, prominent defects related absorption band in the range of 400 to 550 nm is observed in the UV-Vis spectrum (Fig. 4) for all ZnS NPs. Use of the absorption spectroscopy, a highly sensitive to trace quantity impurity [7] is then used to monitor change in intensity of the absorption band due to surface reactivity with $Ag^+$ in solution. $AgNO_3$ is used for $Ag^+$ source for investigation of surfaces reactivity with ZnS NPs.

Typical UV-visible absorption spectra of 3 nm ZnS NPs and after addition of various concentrations of $AgNO_3$ are shown in Fig. 6. Here 1 mg/ml solution of ZnS NPs was used to detect different concentration of $Ag^+$. Distinct broad peak around 400 – 550 nm (Fig. 6) is seen to arise from various defects, predominantly from 'S' dominant region as discussed in PL study (Fig. 5). It is to be noted that decreases in the intensity of absorption spectra of ZnS NPs occur with increasing concentrations of $Ag^+$ (Fig. 6). This alteration is directly proportional to different concentrations of $Ag^+$ and it is favorably used for the detection and estimation of $Ag^+$ (Fig. 7). As far response time concerned it is governed by spectrometer and the interaction time. Spectrum was recorded immediately after addition of $Ag^+$ to the ZnS solution and the interaction takes place in fraction of second. Interestingly, exothermic $Ag_2S$ formation was reported to be very quick even in thin film structure [43]. Spectrometer response is as fast as (20 x 50 ms) where each spectrum is recorded for 20 ms with 50 cycle. This approach allows cost effective and quick determination of a wide range of concentrations starting from nM to mM $Ag^+$. Other fluorescence technique [18] using ZnS was found to be strongly quenched within 0.5-1 µM range.

### 3.3. ZnS NPs size effect on detection

Importantly slight variation in intensity of absorption is observed for different particle sizes of ZnS NPs. Such variation arises possibly from the variation of surface areas due size and the presence of surface related defects. For systematic and comparative data analysis from acquired UV-Vis spectra, the highest observed intensity at 500 nm (2.48 eV) from the broad feature is chosen. This region belongs to dominant interaction sites to 2.3 and 2.55 eV bands as observed in PL measurements (Fig. 5). The variation in absorption intensity from ZnS reference at 500 nm with addition of $Ag^+$ is plotted as semi Log scale in Fig. 7 for all three ZnS NPs. Error bars in Fig. 7 indicates statistical error involved with repeated experiments for the same concentration. A monotonic decrease in absorption intensity is seen with $Ag^+$ concentrations. A straight line is found to fit through the wide range of concentrations (Fig. 7), and the characteristic curve for different particle sizes can be defined by the equation 3.

$$Y = a + b \, Log(X) \quad\text{------------------------------------------------------------- (3)}$$

where X is molar concentration of $Ag^+$ and Y is intensity of absorption. '*a*' is a constant and '*b*' is slope. Importantly, these fitting acts as calibration curve for a typical particle size and can allow finding an unknown concentration. The parameter *a*, a constant, is found to depend on ZnS NPs. The value of slope, *b* defining the variation of absorption with concentration indicates the sensitivity towards $Ag^+$ detection. These parameters for different ZnS NPs are given in Table 2. Notably the sensitivity value is found to be highest for the smallest 3 nm ZnS NPs which offer maximum surface area with largest possible surface defects dominated by 'elemental' type 'S' (Fig. 5). Both factors surface area and defects favor interactions. Attempt is made to understand detection at low concentration also. Absorption peak is found to vary linearly for relatively low concentrations of $Ag^+$ (~ nM to μM range) (inset Fig. 7). However, with increasing $Ag^+$, availability of typical defect sites in ZnS NPs were not be sufficient for competitive interactions in the same way. Subsequently, the absorption peak deviated from the linear behavior at high concentration of $Ag^+$. However, response for $Ag^+$ is highly possible even for mM concentrations (Fig. 7) ensuring widespread detection from nM to mM range. This is to mention that very low spectrometer noise level (less than 0.1%) in absorption scale allows trace amount detection. Stern-Volmer equation given below is used for Ag ion detection by researchers using fluorescence method [14, 21, 24].

$$x = \frac{A_0}{A} = 1 + K_{sv} \cdot [Ag^+] \quad\text{-------------------------------- (4)}$$

Here, $A_0$ and A stand for without addition of $Ag^+$ and in presence of $Ag^+$ whereas $K_{sv}$ refers to the Stern–Volmer constant. In line with above, the observed decay of absorption arising from association and interaction from $Ag^+$ and ZnS QDs is also tested and fitted with the above equation for low concentration regime (~ nM). Notably, the equation is fitted linearly and the corresponding $K_{sv}$ for 3 nm QDs is found to be $4.48 \times 10^6$ L/mol. This value matches closely for similar system [14, 24] indicating static quenching of absorption due to association [44].

### 3.4. Interaction mechanism for detection

To circumvent the speculative interaction between $Ag^+$ and ZnS NPs, further study is carried out. A PL spectrum of ZnS NPs after an addition of $10^{-4}$ M $Ag^+$ was recorded for understanding the role of various defects on the interaction with ZnS (Fig. 8). It shows strong decrease in intensity of defect peaks centered at 2.3 and 2.55 eV. Such strong quenching of PL intensity points to preferential interaction of $Ag^+$ with elemental type S and Zn vacancy sites of ZnS NPs. Change of color of the dispersed ZnS in water was also recorded after and before addition of $Ag^+$ (Fig. 8). One of the tubes displays opaque dispersion of ZnS NPs in water. On the addition of 1 ml of 1μM $AgNO_3$ to 1 ml of ZnS NPs, a clear change in the color was observed (second tube). On further addition to mM, a black precipitate settled in the third tube. The formation of black coloration is an indication of formation of $Ag_2S$. For further investigation, it was washed and centrifuged with water several times before it was dried at 70 °C. Powder XRD measurement was carried out for the black precipitate (Fig. 9). XRD pattern showed broad as well sharp peaks. Broad peaks matched well with cubic ZnS crystal where as sharp peaks corresponded to monoclinic $Ag_2S$ phase (JCPDS # 00-014-0072). This confirms the $Ag_2S$ formation as a result of interaction between $AgNO_3$ solution and ZnS NPs. Furthermore, Williamson – Hall plots (inset of Fig. 9) were drawn to evaluate defect induced strain in the system. A strain of 2.5% was seen in ZnS NPs. However, it is released after the interaction leading to formation of $Ag_2S$ (inset of Fig. 9) supporting participation of defects site for $Ag^+$ detection. The size of $Ag_2S$ NPs deduced from XRD patterns (Fig. 9) is around 20 nm which is bigger in size than that of the ZnS QDs. This observation points to aggregation of $Ag_2S$ after reaction ZnS NPs at room temperature.

### 3.5. Analysis of other metal ions

Similar to above experiment using $NaNO_3$ solution instead of $AgNO_3$ was carried out. In this particular case, the size of Na cation (102 pm) was similar to Ag ion (115 pm). However, no strong change in UV-Vis spectrum was noticed. This result indicated strong specificity of $Ag^+$ towards sulfur binding over the $Na^+$. Reaction of S with $Ag^+$ led to strong covalent bond formation in $Ag_2S$ which is insoluble ($K_{sp}\sim10^{-51}$) in water medium whereas $Na_2S$ being relatively ionic compound is highly soluble [45]. This solubility maintained chemical equilibrium between $Na_2S$ and ZnS with limited precipitate which in turn gave negligible change in absorption for detection (Fig. 10). Additionally, with regard to 'S' prone elements, study was carried out with hazardous metal ion like Pb, Hg and Cd. Fig. 10 displays reactivity of heavy metals ions in aqueous medium in comparison to $Ag^+$ for mM concentration. Responses with statistical error for different metals ion are displayed in Fig. 10. The response is defined as $R = (ZnS_{abs}-X_{abs})/ZnS_{abs}$, where $X_{abs}$ is absorption value at 500 nm after addition of metal nitrate solutions to reference absorption ($ZnS_{abs}$). Maximum change was observed for $Ag^+$ and was considered as 100% response. Comparatively, the response for other metal ions Na, Cd, Pb, and Hg were 7%, 17%, 43% and 73%, respectively. It is clear that metal ions known for 'S' affinity like heavy and toxic metals ions (Cd, Pb, and Hg) with low solubility provide response. They too can be detected separately. It is noteworthy that strongest response for $Ag^+$ in aqueous medium ($K_{sp}$~$10^{-51}$) over other ions. The solubility product for HgS is equivalently high ($K_{sp}$~$10^{-53}$) whereas PbS has $K_{sp}$~$10^{-28}$. However, responses do not scale linearly with solubility factor alone indicating other factors like size, rate constant, and valence state of the metal ions also play crucial role for precipitate and detection. In addition to above, understanding selectivity of $Ag^+$ with reference to commonly present ions in aquatic medium, detection experiment was carried out for $K^+$, $Ca^{2+}$, $Mg^{2+}$, $Mn^{2+}$ and $Ba^{2+}$. They showed (The Inset in Fig. 10) typical response observed for Na salt discussed above. The response for $K^+$, $Ca^{2+}$, $Mg^{2+}$, $Mn^{2+}$ and $Ba^{2+}$ were found to be nearly 7%, 5%, 8%, 8% and 10%, respectively. Thus $Ag^+$ is found to be highly selective among $Na^+$, $K^+$, $Ca^{2+}$, $Mg^{2+}$, $Mn^{2+}$ and $Ba^{2+}$. In the present study, a concept proof method is therefore developed and is found to be a leap forward for usefulness of this well known bio-compatible metal sulfide NPs in aqueous medium in simple cost effective way. This method does require chemical functionality as required for fluorescence, FRET and other methods [7, 14, 16-19, 26]. Moreover, there are several methods of retrieving Ag from $Ag_2S$ by oxidation to Ag and oxides of sulfur [46]. In that case, present uses of inorganic ZnS may be intuitively utilized for retrieval of Ag from $Ag_2S$ and make it environment friendly.

4. **Conclusion:**

ZnS NPs of different particle sizes were synthesized and are utilized for detection of trace amount of $Ag^+$. UV-Visible absorption spectroscopy is utilized for the absorption study and determination of a wide range of concentrations of $Ag^+$. Defect band in ZnS NPs prominently appears at 400 – 550 nm in UV-Vis region and is used for $Ag^+$ detection in aqueous solution. This defect containing surfaces in ZnS allow the formation of insoluble precipitate of $Ag_2S$, which is confirmed using X-ray diffraction measurement. In consequence, total absorption of ZnS NPs is affected and it provides a platform for quantitative estimation for $Ag^+$ detection for a wide range of concentrations starting from nM to mM. Smallest ZnS NPs (3 nm) shows the best performance for $Ag^+$ detection due to significant presence of above surface defects. Above interaction mechanism is further verified and utilized for toxic and heavy elements *e.g.* Pb, Hg and Cd where Ag-S interaction is the most responsive. Moreover, over a large number of commonly present ions ($Na^+$, $K^+$, $Ca^{2+}$, $Mg^{2+}$, $Mn^{2+}$ and $Ba^{2+}$) in aqueous medium were also explored. Among them, best selectivity for $Ag^+$ is found with encouraging and promising utility of the process. This demonstration with simple optical fiber UV Vis spectroscopy makes system easy, cost effective and portable.

## Acknowledgements

One of the author (RJ) would like to acknowledge Shri K V Ravi, CEO, NRB, BARC and Shri M. R. Sankaran, OIC, HPU, NRB, BARCf (K) for their support in performing this work. The authors are grateful to Dr. P. Magudapathy for XRD measurements.

Table 1. Size of ZnS NPs by different techniques.

| Band gap (eV) | Size deduced from band gap (nm) | Size from XRD (nm) | Average size from HRTEM (nm) |
|---|---|---|---|
| 4.5 | 3.15 ± 0.05 | 3.0 ± 0.1 | 3 |
| 4.1 | 3.85 ± 0.05 | 3.5 ± 0.1 | 3.7 |
| 3.6 | - | 4.5 ± 0.1 | 5 |

Table 2. Calculated fitting parameters for different ZnS NPs.

| At 500 nm | 3 nm ZnS | 3.5 nm ZnS | 4.5 nm ZnS |
|---|---|---|---|
| '$a$' value | 0.015 ± 0.005 | 0.069 ± 0.008 | 0.052 ± 0.003 |
| '$b$' value (-ve slope) | 0.014 ± 0.001 | 0.012 ± 0.001 | 0.008 ± 0.0005 |

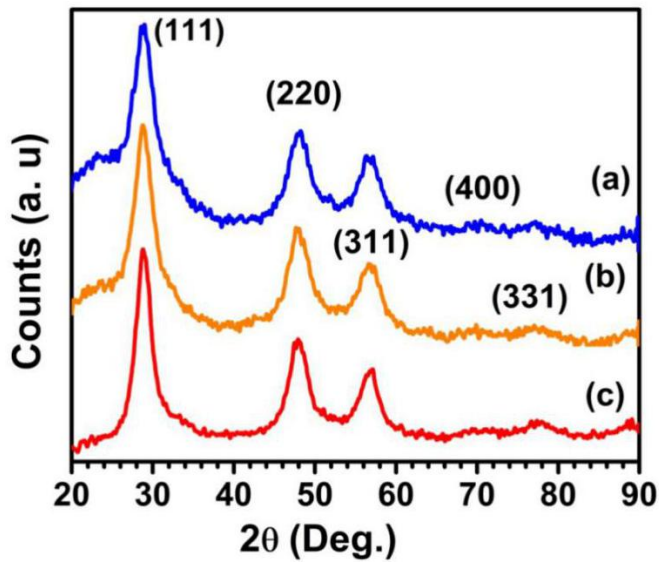

Fig. 1. XRD of various ZnS NPs prepared with concentrations of $ZnCl_2$ a) 0.1 wt% b) 0.25 wt% and c) 0.5 wt%.

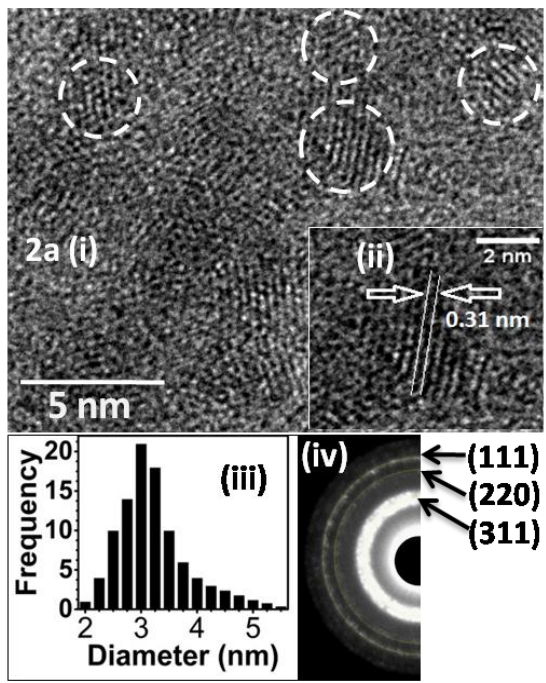

Fig. 2a. HRTEM studies elucidating (i) quantum sized ZnS synthesized from 0.1 wt% $ZnCl_2$, (ii) the *d* spacing of the (111) plane of cubic ZnS phase, (iii) particle size distribution with a maximum around 3 nm and (iv) SAED pattern confirming cubic phase of ZnS.

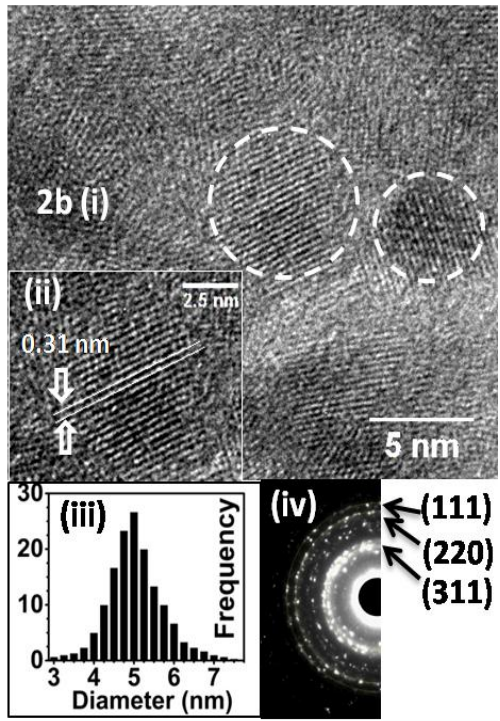

Fig. 2b. HRTEM studies elucidating typical, (i) Quantum sized ZnS synthesized from 0.5 wt% $ZnCl_2$, (ii) the *d* spacing of the (111) plane of cubic ZnS phase, (iii) particle size distribution with a maximum around 5 nm and (iv) SAED pattern confirming cubic phase of ZnS.

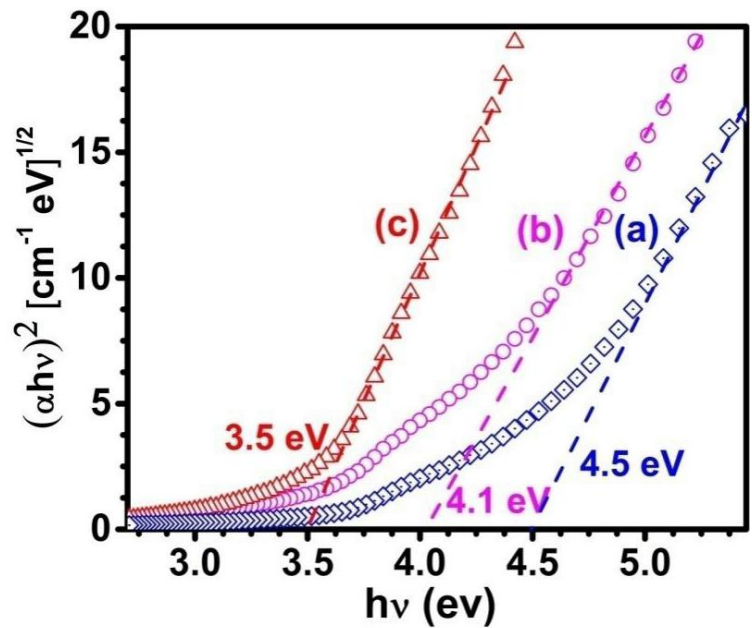

Fig. 3. Tauc's plot for different sizes of ZnS NPs, a) size 3 nm, b) size 3.5 nm, c) size 4.5 nm. Smaller size particle exhibits higher band gap due to quantum confinement effect.

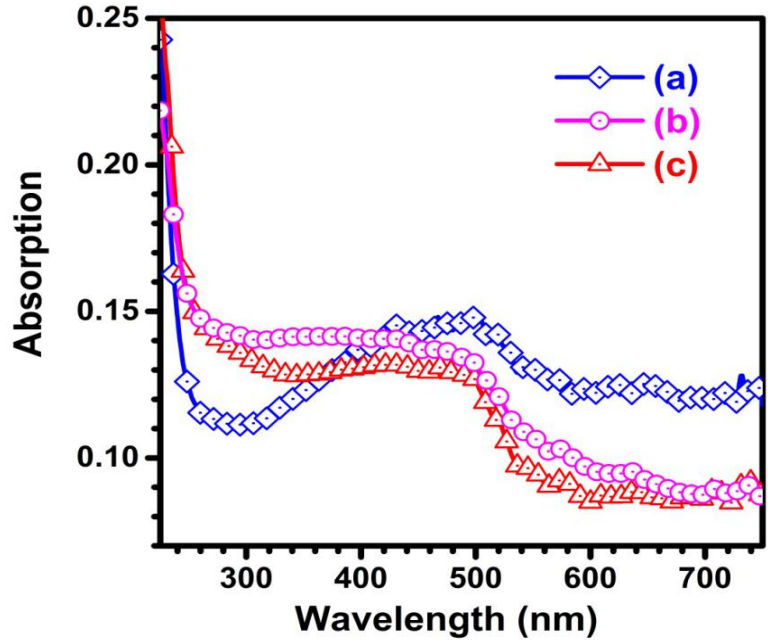

Fig. 4. Comparison of absorption peak arising from defects of three different ZnS NPs a) 3 nm, b) 3.5 nm, c) 4.5 nm. Inset shows variation of absorption at 500 nm with concentration.

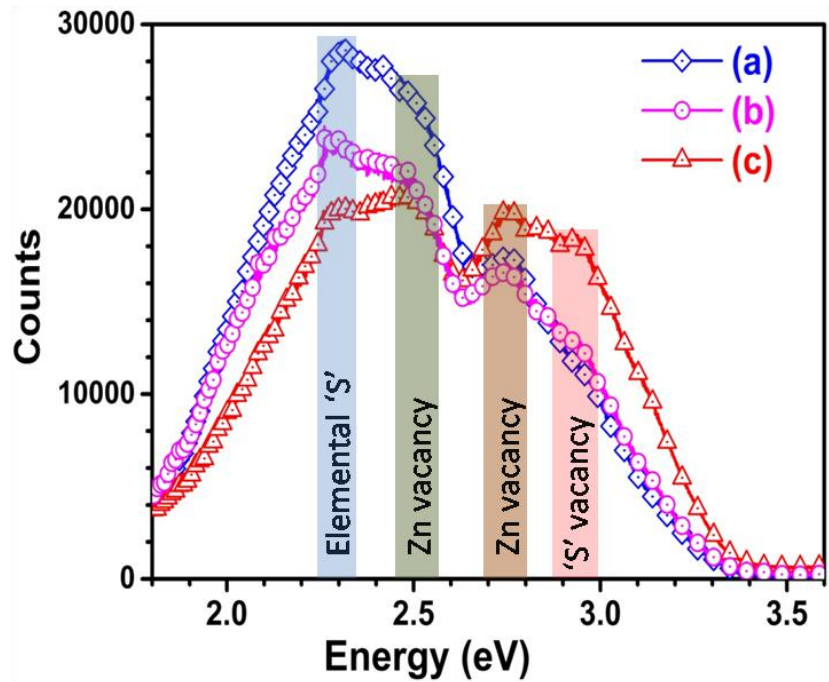

Fig. 5. PL spectra showing presence of defects in ZnS NPs a) 3 nm, b) 3.5 nm, c) 4.5 nm.

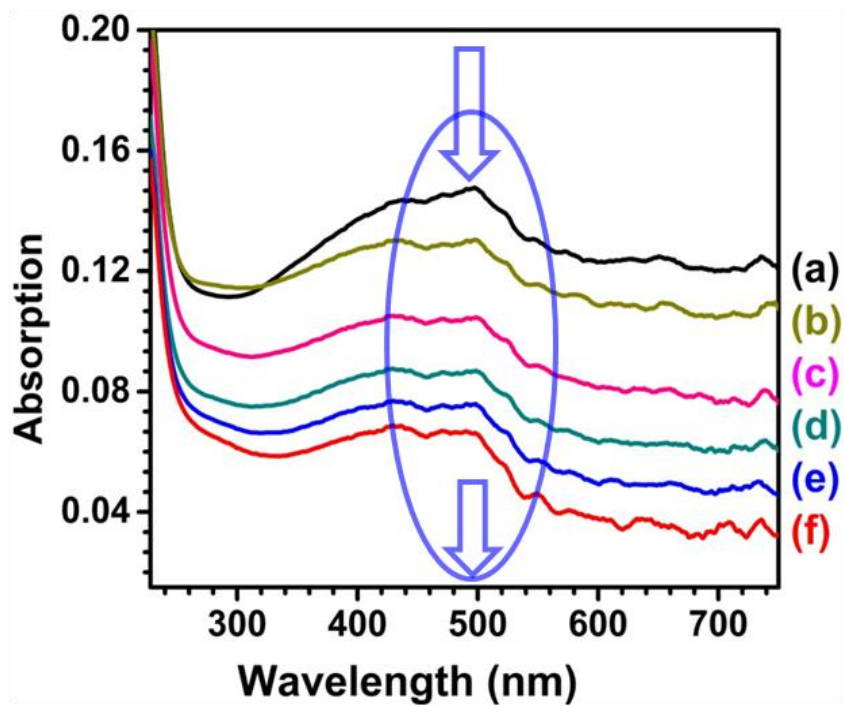

Fig. 6. UV-Vis absorption spectra for a) pure ZnS in water and with the addition of different concentrations of AgNO$_3$ b) 10 nM, c) 0.1 μM, d) 1 μM, e) 50 μM, f) 0.5 mM.

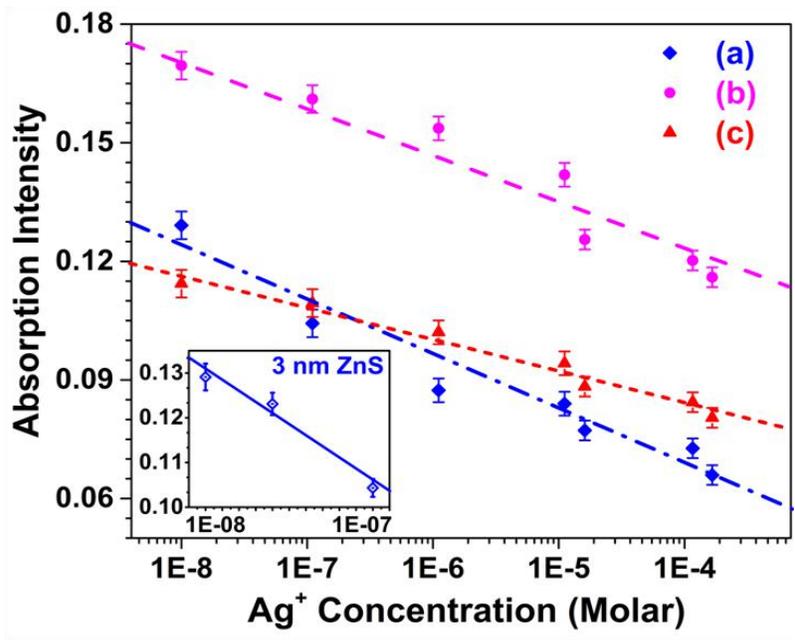

Fig. 7. Optimized response at 500 nm for three different particle sizes. Response curve for a) 3 nm, b) 3.5 nm, c) 4.5 nm ZnS. The inset displays linear variation of response for very low concentrations of Ag$^+$.

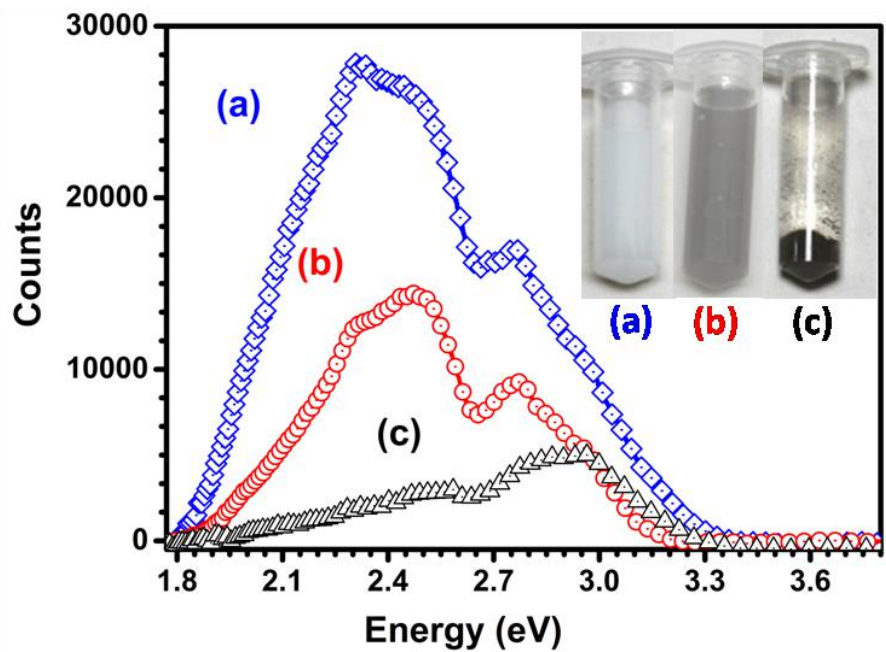

Fig. 8. Typical PL spectra recorded before and after addition of AgNO$_3$ to 3 nm ZnS NPs, a) ZnS NPs, b) with 1 μM AgNO$_3$, c) with 1mM AgNO$_3$. Each tube in the inset displays a) ZnS NPs, b) ZnS NPs with 1 μM AgNO$_3$, c) ZnS NPs with 1 mM AgNO$_3$.

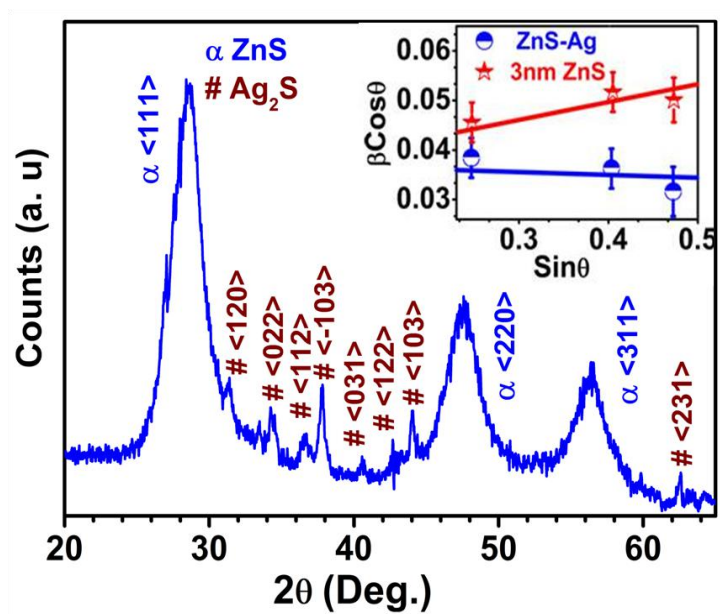

Fig. 9. XRD of ZnS NPs after reaction with AgNO$_3$ solution. The inset is W-H plot for the smallest ZnS NPs (3 nm) and the ZnS NPs after addition of AgNO$_3$.

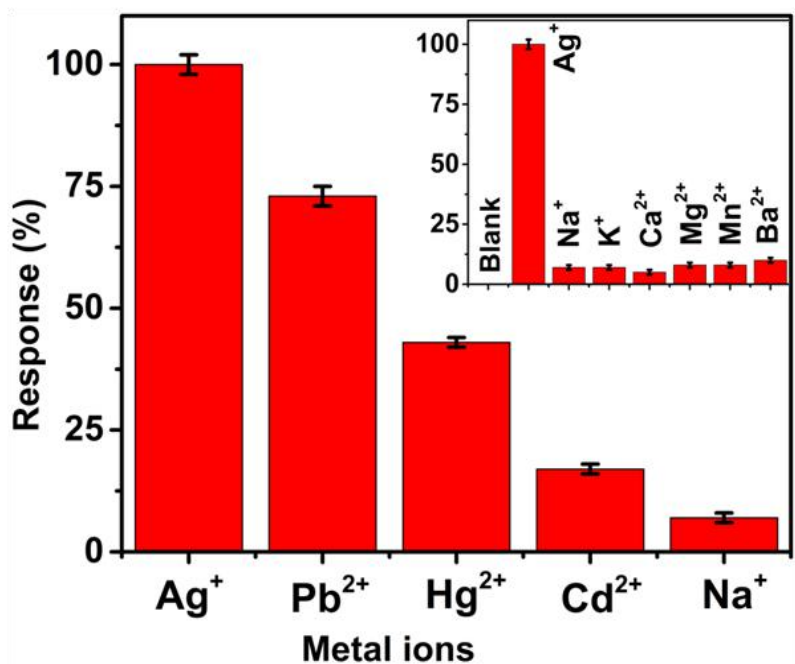

Fig. 10. Response of metal ions to 3 nm ZnS NPs with reference to 100% response for Ag$^+$. The inset displays the selectivity of Ag$^+$ with respect to alkali and alkaline metals which most expected to present in aquatic medium.

**Graphical abstract**

**Nano-molar to milli-molar level Ag (I) determination using ZnS QDs**

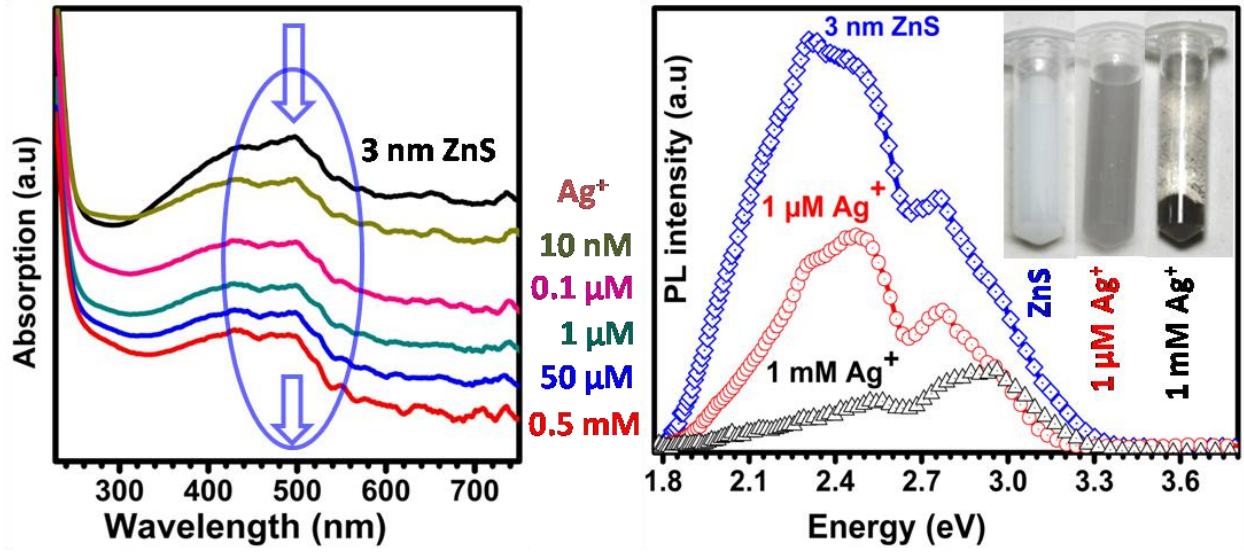